\documentclass[10pt,conference]{IEEEtran}
\IEEEoverridecommandlockouts
\usepackage{cite}
\usepackage{amsmath,amssymb,amsfonts}
\usepackage{algorithmic}
\usepackage{graphicx}
\usepackage{textcomp}
\usepackage{xcolor}
\usepackage{makecell}
\usepackage{multirow}
\usepackage{booktabs}
\usepackage{listings}
\usepackage{caption}
\usepackage[multiple]{footmisc}
\usepackage{url}
\usepackage{verbatim}

\def\BibTeX{{\rm B\kern-.05em{\sc i\kern-.025em b}\kern-.08em
    T\kern-.1667em\lower.7ex\hbox{E}\kern-.125emX}}
\begin{document}

\title{SuMo: A Mutation Testing Strategy for Solidity Smart Contracts}
\pagenumbering{arabic}

\author{\IEEEauthorblockN{Morena Barboni\thanks{Morena Barboni's work to this paper has been partially supported by the Italian MIUR PRIN 2017 Project: SISMA (Contract 201752ENYB)}}
\IEEEauthorblockA{\textit{Istituto di analisi dei sistemi ed informatica (IASI)}\\ Consiglio Nazionale delle Ricerche (CNR) -- Italy \\
morena.barboni@iasi.cnr.it}
\and
\IEEEauthorblockN{Andrea Morichetta}
\IEEEauthorblockA{\textit{School of Science and Technology}\\University of Camerino -- Italy \\
andrea.morichetta@unicam.it}
\and
\IEEEauthorblockN{Andrea Polini}
\IEEEauthorblockA{\textit{School of Science and Technology}\\University of Camerino -- Italy \\
andrea.polini@unicam.it}
}

\maketitle
\thispagestyle{plain}
\pagestyle{plain}

\begin{abstract}
Smart Contracts are software programs that are deployed and executed within a blockchain infrastructure. Due to their immutable nature, directly resulting from the specific characteristics of the deploying infrastructure, smart contracts must be thoroughly tested before their release. Testing is one of the main activities that can help to improve the reliability of a smart contract, so as to possibly prevent considerable loss of valuable assets. It is therefore important to provide the testers with tools that permit them to assess the activity they performed.

Mutation testing is a powerful approach for assessing the fault-detection capability of a test suite. In this paper, we propose SuMo, a novel mutation testing tool for Ethereum Smart Contracts. SuMo implements a set of 44 mutation operators that were designed starting from the latest Solidity documentation, and from well-known mutation testing tools. These allow to simulate a wide variety of faults that can be made by smart contract developers.
The set of operators was designed to limit the generation of stillborn mutants, which slow down the mutation testing process and limit the usability of the tool. We report a first evaluation of SuMo on open-source projects for which test suites were available. The results we got are encouraging, and they suggest that SuMo can effectively help developers to deliver more reliable smart contracts.
\end{abstract}


\section{Introduction}
\label{sec:introduction}

In recent years there has been an increasing focus on using blockchain in the engineering of software systems  \cite{Porru2017,sac20-chorchain}. Compared to traditional software, smart contracts development presents unique challenges due to the underlying blockchain environment \cite{Zheng2017, Destefanis2018}. These underline the need for appropriate testing tools that allow the developer community to write and deploy safer code. The following non-exhaustive list defines the mains reasons as to why smart contracts ask for high-reliability guarantees, and a thorough testing process.
%\begin{itemize}
    %\item 
    \textbf{Smart Contracts manage valuable assets}: Smart contracts can control large amounts of cryptocurrency and other valuable assets. Deploying faulty code can result in the accidental loss of the assets held by the contract. The potential financial gain and the anonymous nature of the blockchain further act as an incentive for attackers. Even a small loophole in the code can allow malicious users to drain large amounts of funds. A typical example is the famous DAO attack \cite{Mehar2019}, in which a reentrancy vulnerability was exploited to transfer 3.6 million Ether (around \$50 million)
    %\item 
    \textbf{Transactions are irreversible}: Smart contracts are deployed and executed in the blockchain environment, which does not allow to revert transactions. A transaction becomes irreversible once it receives enough confirmations from the network. At the same time, it is not possible to recover assets lost during the smart contract execution.
    %\item 
    \textbf{Smart Contracts are immutable}: Smart contracts feature an anomalous development life cycle \cite{Miraz2020} that cannot be represented by traditional software development models. Enhancing the code or fixing bugs after deployment is not possible. Immutability ensures that the code is tamper-proof, but it also prevents further upgrading and code maintenance. Correcting a bug after deployment is a very costly operation since it requires the creation of a brand new contract on the blockchain.
    %\item
    \textbf{Blockchain environment}: The smart contract execution is dependant on the underlying blockchain platform and the possible interactions with other cooperating contracts. Developers must carefully consider these relationships and the peculiarities of the new distributed environment to write safe code. 
    %\item
    \textbf{New software stack}: The smart contract’s execution environments and programming languages (e.g. Solidity) are relatively young and continuously evolving; many issues and vulnerabilities are still being discovered. Such characteristics make it harder for developers to program with confidence and to write safe code. 
    % \item 
    \textbf{Lack of best practices}: Zou et al.'s \cite{Zou2019} recent interview with smart contract experts exposed a lack of best practices for writing reliable contract code.  Finding code examples and development standards is particularly difficult when working on new applications. This issue can cause developers to pick up bad programming habits and dangerous anti-patterns. An anti-pattern is a solution created in response to a recurring problem, which appears to be appropriate and effective. However, it ends up being ineffective or even counterproductive.
     %\item 
     \textbf{Lack of mature testing tools}: Smart contract development cannot count on the wide selection of testing tools available for traditional software. The currently available tools are not as mature, and in some cases, they are ineffective at ensuring the quality of the contract code. Zou et al.’s \cite{Zou2019} research shows that the developer community is especially interested in code auditing tools, which help in discovering bugs and vulnerabilities.
%\end{itemize}

The testing activity allows deriving confidence in the correctness and reliability of a smart contract. However, this level of confidence is strictly dependent on the adequacy of the implemented test suite. Most white-box adequacy criteria are defined based on code coverage, which measures how thoroughly the logical elements within the software are exercised by the test suite. If certain elements are not covered by tests, they might contain faults that persist in the system after testing. Even though code coverage helps in identifying under-tested parts of a program, research shows that it does not represent a good indicator of the test suite effectiveness \cite{Inozemtseva2014,Tengeri2016}. Besides, it was proven that more complex forms of coverage do not provide much greater insights into the effectiveness of the test suite.

In this paper, we propose a novel mutation testing strategy tailored for smart contracts to be deployed on the Ethereum blockchain. The strategy has been implemented in SuMo (that is a mutated version of SoMu, which stands for \textbf{So}lidity \textbf{Mu}tator) and applied to assess the test suites of two open-source and published smart contracts.

The rest of the paper is organized as follows. In the next section, we include some background material. Section \ref{sec:methodology} details the SuMo strategy and the included mutation operators, while in Section \ref{sec:validation} we illustrate an initial validation activity we performed. Section \ref{sec:related} reports those works that are closer to our proposal. Finally Section \ref{sec:conclusion} describes some conclusions, and opportunities for future work.

\section{Background}
\label{sec:background}
\subsection{Ethereum Smart Contracts}
A smart contract is a program that runs on top of a blockchain network. Smart contracts are commonly referred to as ‘self-enforcing agreements’ because the execution process automatically enforces the terms defined within the code. This property allows the execution of transactions among untrusted parties without the need for a central authority. Each contract can implement complex business logic to carry out transactions that go beyond a simple monetary transfer. \textbf{Ethereum} is a blockchain technology specifically built for the creation and execution of smart contracts. In particular, Ethereum introduced support for Turing-complete programming languages, such as Solidity, within a blockchain infrastructure.

\subsubsection{Ethereum Accounts}
The Ethereum platform is based on the account model. Each account is an object identified by a 20-byte address and associated with a state. There are two types of accounts:    (1) \textbf{Externally Owned Account (EOA)}: is controlled by a private key. The owner of the key can perform transactions and manage the funds associated with the account. (2) \textbf{Contract Account (CA)}: is controlled by the code. When triggered by a transaction, a CA can perform operations on its storage and fire calls to other contracts. The \textbf{Account State} comprises the following information: 
    \textbf{nonce}: a counter that stores either the number of transactions executed from an EOA or the number of generated contract instances for a CA;
    \textbf{balance}: the amount of Ether held by the account;
    \textbf{storageRoot}: the Merkle-Patricia trie root that encodes the storage contents of the account; It is only relevant for a CA.
    \textbf{codeHash}: hash of the code associated with a CA.

\subsubsection{Ethereum World State}
Ethereum was defined in the Ethereum Yellow Paper \cite{wood2019ethereum} as a \textbf{transaction-based state machine}. Unlike Bitcoin, Ethereum has a global state that enables the management of real account balances. Each Account State is stored inside a \textbf{World State trie} data structure. Every time transactions take place, the global state of the blockchain is updated. The Ethereum blockchain can be seen as a state transition system, where a state transition takes a state and a transaction and outputs a new state as a result. The last block of the chain represents the current world state. This block holds information about the balances of all the Ethereum accounts after the most recent transactions have been executed.

\subsubsection{Ethereum Transactions}
A transaction is a digitally signed message issued by an EOA. 
Three types of transactions can be executed with different purposes:
    \textbf{Monetary}: A monetary transaction is a simple transfer of funds between EOAs.
    \textbf{Contract Deployment}: A contract deployment transaction allows the sender to create a CA on the blockchain. The transaction must contain the compiled contract code as a payload. 
    \textbf{Contract Execution}: A contract execution transaction runs a function defined within a deployed contract. 
    Upon execution, the smart contract can perform operations on its own state and send message calls to other CAs. 

The execution and confirmation of a transaction require several steps. First, the sender must create a transaction and broadcast it to the Ethereum network. The miner nodes are responsible for including new transactions into a \textbf{candidate block}. Each miner selects the transactions to include in the block from a \textbf{transaction pool} and starts validating them. This process involves checking the digital signature and the output of any contract code execution. Once the candidate block is built, the miner starts working to find a valid Proof of Work solution. When a valid hash is found the candidate block is broadcast to the rest of the network. Upon reception of the block, each full node repeats the validation process for each transaction. The nodes do not compare the results among themselves, but they rely solely on the \textbf{consensus mechanism}; if at least one transaction is invalid every honest node will reject the block. A valid block cannot be considered part of the main chain until it is confirmed. The distributed nature of the blockchain can cause the nodes to receive blocks in a different order. When separate parts of the network follow distinct solutions the main chain \textbf{forks}. Ethereum selects the only valid branch using the \textbf{GHOST (Greedy Heaviest Observed Subtree)} protocol \cite{ethereumWhitepaper}. If a valid block is part of a discarded fork, all the included transactions go back to the transaction pool. Otherwise, the transactions will be permanently committed to the blockchain. In this case, users must wait for a confirmation which is generally issued after 12 blocks.

\subsubsection{Gas Mechanism}
Each node of the Ethereum network includes the \textbf{Ethereum Virtual Machine (EVM)}, a stack-based execution environment for running smart contract code. The EVM allows each Ethereum node to participate in the verification protocol. Since each contract is locally executed by the whole network the validation process can become computationally expensive. Moreover, non-terminating program executions could potentially freeze the participating nodes. Since the EVM supports Turing complete languages, predicting the amount of required computational resources is not simple. Ethereum implements a \textbf{gas mechanism} that serves two main purposes: It limits the usage of resources, and it compensates miners for their work. The issuer of a transaction must pay a gas fee to the client that commits the transaction to the blockchain. This fee is determined by the amount of computation and data storage required by the transaction execution. Each transaction object contains a \texttt{gasLimit} and a \texttt{gasPrice} field. The \textbf{gas price} indicates the market price in Wei of a unit of gas. The \textbf{gas limit} is the maximum amount of gas that can be burnt for executing the transaction. The amount of gas needed for executing a smart contract depends on the number and type of instructions run by the EVM. 

\subsubsection{Solidity}
Solidity is the most widely used object-oriented language for developing smart contracts. When the high-level Solidity code is compiled to bytecode, it can be run on the EVM to enforce the terms described within the contract. The syntax of Solidity is mainly influenced by JavaScript, but it is statically and strongly typed. Solidity supports inheritance, polymorphism, and user-defined data types. Moreover, it introduces several novel constructs and keywords dedicated to smart contracts development.

\subsection{Mutation testing} 
\label{sec:mutation}
Mutation testing is a white-box, fault-based testing technique for evaluating and enhancing the quality of a test suite. Mutation testing aims to guarantee the adequacy of test data in finding real faults by purposefully introducing small defects into the source code. This process helps to identify limitations in the implemented test suite and to ensure the deployment of more reliable code. However, generating mutants that simulate all the possible faults that can be found in a real-world program is not possible. For this reason, this technique only targets simple faults that are somehow close to programming errors. Mutation testing is based on the assumption that a subset of simple faults is sufficient to simulate the totality of faults.
The traditional mutation process generates a set of faulty copies of the original program P called \textbf{mutants}. A mutant does not deviate largely from the original program, but it contains a minor alteration that simulates a typical programming mistake. The core element of mutation testing is a set of rules called \textbf{mutation operators}. Each operator specifies how the source code must be altered to create a mutant. The adequacy assessment is performed by running the test suite T against each mutant program, and verifying whether the tests can detect the fault injected into the code. If at least one test case in T can detect the fault, the mutant is said to be killed. The percentage of killed mutants, called \textbf{Mutation Score} (Equation \ref{eqn:mutation.score}), represents the adequacy value of the test suite.
\begin{equation}
\label{eqn:mutation.score}\footnotesize
    MS = \frac{Non\,equivalent\,mutants  - surviving\, mutants}{Non\,equivalent\,mutants}\times 100
\end{equation}

The developer can iteratively improve the test suite until the adequacy criterion is satisfied by the mutation score. However, before calculating the mutation score, it is necessary to identify and remove the equivalent mutants.  An \textbf{equivalent mutant} is a program that contains syntactic modifications, but it behaves exactly like the original program.
\begin{comment}
Undetected equivalent mutants decrease the reliability of the mutation score because they are mislabeled as faulty programs. The analysis of the alive non-equivalent mutants can provide useful insights about the types of faults that cannot be caught by the existing tests. Based on this feedback it is possible to enhance the test suite and cover scenarios that were neglected during the testing design phase. 
\end{comment}

\section{Mutation Strategy}
\label{sec:methodology}

%\subsection{Challenges}
One of the main challenges of implementing a mutation testing strategy is designing a comprehensive set of mutation operators that can simulate a wide range of faults. In fact, a test suite can be improved based on the introduced faults that it is not able to detect. A strategy based on a limited fault model cannot ensure the derivation of a high-quality test suite. To build the fault model of SuMo we carried out an in-depth analysis of the Solidity language\cite{SolidityDoc} and of existing vulnerability taxonomies \cite{Atzei2017, Dika2018}.  This analysis aimed at identifying opportunities for designing new mutation operators (or improving the existing ones). Meaningful mutation operators should simulate likely faults that can manifest as anomalous contract behavior. The mutation should also be reasonably reachable from the test suite. Therefore, we focused on potential errors in the programmer's thought process and their effects on the contract execution. The second challenge in the design of SuMo was guaranteeing good usability and effectiveness. SuMo attempts to alleviate the cost of the mutation process by applying different strategies.

%\begin{itemize}
    %\item 
    \textbf{Customized Mutation Process:} SuMo allows the tester to select which mutation operators to apply and which contracts to mutate. By default, starting a mutation process runs all the implemented mutation operators. However, depending on the test suite and the contract under test, the developer might only want to apply specific mutations. Disabling selected operators can significantly reduce the number of generated mutants, and speed up the mutation testing process.\\
    %\item 
    \textbf{Limiting Redundant Mutants:} A redundant mutant generates an outcome that can be predicted based on other mutants. SuMo attempts to limit the generation of redundant mutants without sacrificing the quality of the test suite. This is achieved by merging those mutations that are more likely to generate redundant mutants in the scope of each operator.
    %\item 
    \textbf{Limiting Stillborn Mutants:} Stillborn mutants represent an unnecessary waste of CPU resources and contribute to slowing down the mutation testing process. Stillborn mutants can be generated when a mutation operator injects a semantic error into the code.  Even if the mutant complies with the syntax of the language, it might still fail during compilation. SuMo tackles this challenge on different levels. First of all, SuMo excludes those operators that generate large amounts of stillborn mutants, such as the ones that target data types or the inheritance mechanisms. In fact, they cause a performance overhead while rarely providing useful information about the test suite. Now, some of the implemented operators can still generate a varying percentage of stillborn mutants. This issue is addressed by implementing precondition checks during the mutation process. In particular, SuMo collects semantic information during the visit of the AST to determine whether a mutation should be applied or not. However, performing multiple visits of the tree is a costly operation that slows down the mutant generation process. Because of this SuMo tolerates the generation of a relatively small number of stillborn mutants.
%\end{itemize}

As a result SuMo currently implements a set of 25 Solidity-specific operators and 19 general operators. These are illustrated in table \ref{tab:sumo.solidity.operators}, where we highlighted in bold those mutation operators that either target features that were never considered before or apply updated mutation strategies. In the following subsections, we present the two categories of operators separately. For the sake of space, we focus the presentation more on the operators we have conceived, and on those for which we included relevant modifications. A comprehensive discussion on all implemented operators can be found on the SuMo page (\url{http://pros.unicam.it/sumo}).

\begin{table}
\caption{SuMo mutation operators}
\label{tab:sumo.solidity.operators}
\centering
\begin{tabular}{ll}
\toprule
\textbf{Operator} & \textbf{Name} \\
\midrule
AVR & Address Value Replacement \\
\textbf{CCD} & \textbf{Contract Constructor Deletion} \\
DLR & Data Location keyword Replacement \\
DOD & Delete Operator Deletion \\
EED & Event Emission Deletion \\
EHC & Exception Handling statement Change \\
ETR & Ether Transfer function Replacement \\
FVR & Function Visibility Replacement \\
\textbf{GVR} & \textbf{Global Variable Replacement} \\
\textbf{MCR} & \textbf{Mathematical and Cryptographic function Replacement} \\
\textbf{MOC} & \textbf{Modifiers Order Change} \\
MOD & Modifier Deletion \\
MOI & Modifier Insertion \\
MOR & Modifier Replacement \\
\textbf{OMD} & \textbf{Overridden Modifier Deletion} \\
PKD & Payable Keyword  Deletion \\
RSD & Return Statement Deletion \\
\textbf{RVS} & \textbf{Return Values Swap} \\
SCEC & Switch Call Expression Casting \\
SFD & Selfdestruct Function Deletion \\
SFI & Selfdestruct Function Insertion \\
\textbf{SFR} & \textbf{SafeMath Function Replacement} \\
TOR & Transaction Origin Replacement \\
VUR & Variable Unit Replacement \\
VVR & Variable Visibility Replacement \\
\hline
\textbf{ACM} & \textbf{Argument Change of overloaded Method call} \\
AOR & Assignment Operator Replacement \\
BCRD & Break and Continue Replacement and Deletion \\
BLR & Boolean Literal Replacement \\
BOR & Binary Operator Replacement \\
CBD & Catch Block Deletion \\
CSC & Conditional Statement Change \\
\textbf{ECS} & \textbf{Explicit Conversion to Smaller type} \\
\textbf{ER} & \textbf{Enum Replacement} \\
HLR & Hexadecimal Literal Replacement \\
ICM & Increments Mirror \\
ILR & Integer Literal Replacement \\
LSC & Loop Statement Change \\
\textbf{OLFD} & \textbf{Overloaded Function Deletion} \\
ORFD & Overridden Function Deletion \\
SKD & Super Keyword Deletion \\
SKI & Super Keyword Insertion \\
SLR & String Literal Replacement \\
UORD & Unary Operator Replacement and Deletion \\
\bottomrule
\end{tabular}
\vspace{-0.6cm}
\end{table}

\subsection{Solidity specific  Operators}
\label{sec:solidityop}
To identify mutation operators specifically targeting Solidity characteristics we considered 14 aspects of the language that could help in classifying and identifying useful operators. As said this leads to the identification of 7 new Solidity mutation operators, while the remaining operators find some correspondence in existing works \cite{Andesta2019, Chapman2019, Honig2019, Hartel2019, Wu2019} even though many of them have been reconsidered and slightly modified. SuMo also includes those operators that have been demonstrated to be effective, while mutation operators that generate an elevated number of invalid mutants were excluded from the set. For instance, operators that target the inheritance mechanism might cause a child contract to show unexpected behaviors. However, altering the inheritance hierarchy prevents the contract from compiling most of the time. Mutation operators proposed in the literature that are no longer valid were either removed or updated to comply with the latest Solidity syntax. Several operators were also enhanced to simulate a wider range of faults. When possible, redundant mutations were merged to reduce the total number of generated mutants. The following subsections provide an analysis of the proposed Solidity-specific mutation operators, and then implemented in the SuMo toolset. 
\subsubsection{Function Mutation Operators}
\label{function.operators}
A function is a unit of code that can be executed within a contract as a result of a transaction or a message call. Solidity provides several standard modifiers that can be added to a function to alter its behavior. In Solidity, it is possible to return data from a function using two different syntaxes. Moreover, each function can have multiple return values. These peculiarities can be a source of confusion, and lead to mistakes in the implementation of a function. \textbf{RVS} (Return Values Swap) is a novel operator included in SuMo.
In case a function returns multiple values, the RVS operator swaps them among themselves. RVS targets both types of return structure supported by Solidity. In particular, each return value is swapped with a single compatible return value within the return statement. This allows to limit the number of generated mutants in case of lengthy return structures. Two values are only considered compatible if they belong to the same data type. This allows to prevent the generation of stillborn mutants and speed up the mutation process. Additional operators belonging to such a class are: \textbf{PKD} (Payable Keyword Deletion), and \textbf{RSD} (Return Statement Deletion).

\subsubsection{Visibility Mutation Operators}
Solidity allows to attach visibility modifiers to functions and variables. These alter how the function (or variable) can be accessed by other contracts. 
Ensuring the correct usage of visibility keywords is essential to avoid the deployment of faulty code. Using a visibility keyword that is too restrictive might cause integration issues with other contracts. 
This type of fault does not necessarily lead to a software failure. However, depending on the internal logic of the affected function, an external party could be able to make unauthorized or unintended state changes. For instance, the affected function could perform critical operations such as withdrawing funds (\textit{SWC-105} -- SWC-n can be accessed at \url{https://swcregistry.io/docs/SWC-n}) or self-destructing the contract (\textit{SWC-106}).
State variables are permanently stored in the blockchain to maintain the current state of the contract. Functions can perform operations on these variables to read or update the contract state. 
SuMo implements two mutation operators that simulate faults concerning both function and variable visibility. The \textbf{FVR} (Function Visibility Replacement) operator replaces a function visibility keyword with a different one. There are some exceptions to this rule that we introduced to limit the possible generation of stillborn mutants. FVR does not mutate the receive Ether function and the fallback function, as they can only be declared as external. Moreover, it does not change the visibility of payable functions to internal or private. Lastly, FVR does not delete the visibility keyword, as the function default visibility is no longer allowed. The \textbf{VVR} (Variable Visibility Replacement) operator replaces the visibility keyword of a state variable with a different one. It also adds a visibility keyword to variables with default visibility.

\subsubsection{Function Modifiers Mutation Operators}
\label{function.modifiers.operators}
Solidity supports the definition of modifiers for altering the behavior of functions. When a modifier is attached to a function declaration it can have different effects on its execution. Function modifiers are commonly used for implementing access control mechanisms. 
However, a modifier can also define additional logic that ``extends'' the behavior of the decorated function. Errors in the usage of modifiers can introduce faults of varying severity into the contract. For instance, attaching the wrong modifier to a function might apply unnecessary restrictions to the transaction execution. This prevents the function from running every time the modifier conditions are not respected (\textit{SWC-123}). Similarly, omitting a modifier might prevent a precondition check from taking place. In the worst-case scenario, this will lead to the unintended or malicious execution of critical operations, such as withdrawing funds or destroying the contract. The \textbf{MOD} (Modifier Deletion), \textbf{MOI} (Modifier Insertion) and \textbf{MOR} (Modifier Replacement) operators implemented by SuMo simulate errors concerning the wrong usage of modifiers. These operators were inspired by \textit{Deviant} \cite{Chapman2019} and \textit{ContractMut}\cite{Hartel2019}. \textbf{MOC} (Modifier Order Change) is a novel operator that simulates the erroneous usage of multiple modifiers. 
If a function signature is attached with multiple modifiers, the MOC operator alters their order. As a result, the modifiers will be run in a different order upon the execution of the function. This mutation can have different effects on the contract, such as the function becoming unreachable. 
The \textbf{OMD} (Overridden Modifier Deletion) is a novel operator that targets the modifier overriding mechanism. 
The operator removes an overridden modifier from a derived contract. When the function decorated with the deleted modifier is called, it will be forced to use a modifier of a higher level in the inheritance hierarchy. Executing a different modifier version than the one intended can cause the contract to behave incorrectly. Moreover, the base modifier might apply different restrictions from the overridden one.

\subsubsection{Constructor Mutation Operators}
A constructor is an optional function that is run upon contract creation. The constructor allows initializing the state variables of a contract before the code is deployed on the blockchain. SuMo implements a novel \textbf{CCD} (Contract Constructor Deletion) mutation operator that targets the contract constructor. This operator was inspired by the JDC operator implemented by muJava \cite{Ma2005, Ma2014}, which forces a class to execute the default constructor. The strategy described by Andesta et al. \cite{Andesta2019} is the only one that proposes dedicated mutation operators for altering a contract constructor. However, from Solidity version 0.4.22 the constructor is defined with the new \texttt{constructor} keyword, which requires the definition of a novel mutation. 
The {\bf CCD} operator removes the constructor from a contract by commenting it out. As a result, the user-defined constructor will not be executed before deployment, and the contract state will not be correctly initialized. This mutation allows determining whether the tests ensure the correctness of the contract state after deployment.

\subsubsection{Events Mutation Operators}
\label{event.operators}
Solidity contracts can leverage the EVM logging facilities by firing events. Whenever an event is emitted its arguments are memorized in the transaction log, a special data structure that resides in the blockchain. Events are fired from a contract so that external parties can catch them and trigger a response. Even though the log is not accessible to contracts, events are often used within \textit{DApps} (Decentralized Applications) to monitor the smart contract behavior and trigger code execution. The \textbf{EED} (Event Emission Deletion) operator implemented by SuMo comments out an event emission statement to prevent its execution. This mutation encourages the definition of test cases that include conditions on events. Mutations that replace event invocations are omitted because they can generate many redundant mutants. Since events often log the arguments of the function from which they are emitted, swapping events also has a high chance of generating a stillborn mutant.

\subsubsection{Variable Units Mutation Operators}
Solidity provides two variable units: Ether Units and Time Units. A literal number can take a suffix of \texttt{wei}, \texttt{gwei} or \texttt{ether} to specify a sub-denomination of Ether. Confusing the Ether units can have many dangerous side effects, such as transferring the wrong amount of funds. Using the wrong time unit can cause anomalous contract behavior as well. If the affected value is a trigger for performing actions, this fault can cause a piece of code not to execute, or to execute at the wrong time. SuMo implements the \textbf{VUR} (Variable Unit Replacement) mutation operator to simulate the incorrect usage of variable units.

\subsubsection{Block and Transaction Properties Mutation Operators}
\label{block.properties.operators}

Solidity provides many special variables and functions that allow the retrieval of information about the blockchain. Confusing the available global variables can have many unexpected effects on the contract executions. Moreover, the improper usage of certain global variables, such as \texttt{block.timestamp}, can introduce dangerous dependencies into the contract. Variables that are subject to the miner’s influence should never be used as a source of entropy (\textit{SWC-120}), or to trigger time-dependent events  (\textit{SWC-116}). 
The novel \textbf{GVR} (Global Variable Replacement) operator implemented by SuMo is designed to avoid the incorrect usage of the global variables and functions available in Solidity. It can also prevent the developer from inserting dangerous dependencies into the contract.
The replacement rules were designed to limit the generation of stillborn mutants as well as the total number of mutants. GVR only performs replacements of variables that represent likely programming mistakes, and that return a compatible data type.
\textbf{TOR} is similar to GVR, but it targets a subset of global variables for retrieving the sender of a transaction. 

\subsubsection{Global Functions Mutation Operators}
Solidity provides many global functions for performing different types of operations. These include the \texttt{addmod()} and \texttt{mulmod()} mathematical functions, which are used for modulo addition and multiplication respectively. Since they take in the same parameters, a developer could easily confuse them without noticing the error. Using the incorrect function leads to the computation of wrong values, as well as potential overflow conditions. Solidity also provides the \texttt{keccak256()}, \texttt{sha256()} and \texttt{ripemd160()} global cryptographic functions. These can be used for a variety of tasks, including hash calculation and public key retrieval. Lastly, Solidity provides a special \texttt{selfdestruct} function to remove a smart contract from the blockchain. When executed, the \texttt{selfdestruct(address payable} \texttt{recipient)} method sends all the Ether stored in the contract to the designated address. The contract’s data is then cleared from the state. The incorrect implementation of \texttt{selfdestruct} can be extremely dangerous, as there is no way of recovering the data or the Ether held by a destroyed contract. SuMo implements several operators that target global functions. The \textbf{MCR} (Mathematical and Cryptographic function Replacement) is a novel operator that helps to spot mistakes in the usage of mathematical and cryptographic functions. In particular, it replaces a mathematical or cryptographic function with a different function of the same type. Other operators in this class are \textbf{SFD} (Selfdestruct Function Deletion), and \textbf{SFI} (Selfdestruct Function Insertion).

\subsubsection{Address Mutation Operators}
Solidity provides a special \texttt{address} type for handling contracts. Since addresses are required for performing a wide variety of operations, an address error can propagate in many unexpected ways.  More in general, a call to the wrong address will either target a different contract than intended or a non-existent contract. If the faulty address exists, the invoked function will likely not match any identifier of the recipient contract. This will either cause the transaction to fail, or the recipient fallback function to execute (\textit{Call To the Unknown}). If the target address is orphaned, the transaction to the non-existent contract will fail. This is not a big issue in the case of normal function calls. However, sending Ether to an orphaned address results in the permanent loss of funds.
SuMo implements several operators that target contract addresses and their members. \textbf{AVR} (Address Value Replacement) helps in avoiding mistakes regarding the usage of addresses. The \textbf{SCEC} (Switch Call Expression Casting) operator was imported from Deviant for ensuring the correct casting of address types.

\subsubsection{Ether Transfer Mutation Operators}
\label{ether.transfer.operators}

Solidity provides several methods for transferring Ether between contracts: \texttt{transfer()}, \texttt{send()}, \texttt{call()}, \texttt{staticcall()} and \texttt{delegatecall()}.
They differentiate among each other concerning the gas limit and the returned value in case of error. The \textbf{ETR} (Ether Transfer function Replacement) operator replaces an ether transfer function with a different one. ETR is similar to operators designed for other proposals. However, in SuMo it was adapted to target the new syntax for the \texttt{call()} method, and it was augmented to  mutate the \texttt{delegatecall()} and \texttt{staticcall()} methods.

\subsubsection{Data Location Mutation Operators}
Every reference type in Solidity is associated a data location keyword that specifies where the variable is stored, where \texttt{memory} is used for storing temporary variables and \texttt{storage} is used for persistently storing state variables.
Confusing the data location keywords alters the behavior of the affected variable. Replacing \texttt{storage} with \texttt{memory} limits the lifetime of the variable to the execution of the function in which it is declared. Conversely, swapping \texttt{memory} with \texttt{storage} extends the lifetime of the variable. The gas cost is affected as well because storage variables are more expensive than memory variables. The \textbf{DLR} (Data Location Replacement)  operator swaps the \texttt{memory} and \texttt{storage} data location keywords.

\subsubsection{Delete Mutation Operator}
The \texttt{delete} keyword is a Solidity-specific operator that assigns the default value to a variable. SuMo implements the DOD mutation operator which was imported by \textit{MuSC}.
This removes any instance of the \texttt{delete} keyword from a contract, one at a time.

\subsubsection{Exception Handling Mutation Operators} Solidity manages exceptions with three different state-reverting functions. Since transactions are atomic, reverting all changes is the only way of preventing unsafe contract executions. When triggered, these functions revert all modifications to the state in the current call, including all sub-calls, and flag an error to the caller. 
%\begin{itemize}
    %\item 
    \texttt{require()} is used to validate conditions that can only be checked at run-time. If the condition is not met, an exception is thrown and the unused gas is refunded to the sender.
    %\item 
    \texttt{assert()} is only used for invariants and internal error checking. Upon error the transaction fails and the state changes are reverted without refunding gas to the sender. 
     %\item 
     \texttt{revert()} is similar to require but it is used for handling business logic errors.
%\end{itemize}
Errors in the usage of exception-handling statements can have different effects. \texttt{require()} and \texttt{revert()} should not be used for internal error checking, while \texttt{assert()} should not be used for checking external inputs (\textit{SWC-110}). A buggy or missing \texttt{assert()} statement does not ensure the proper validation of the contract. Likewise, if a \texttt{require()} or \texttt{revert()} statement contains a fault, the run-time conditions will not be properly checked. 
The \textbf{EHC} (Exception Handling statement Change) operator removes any instance of exception handling statement from a contract, one at a time. This mutation  prevents the condition check from taking place. It also replaces an instance of exception handling statement with a different one. In particular, the run-time \texttt{require()} statement is swapped with \texttt{assert()} and vice versa. This mutation simulates the usage of an exception handling statement in the wrong context. 

\subsubsection{Libraries Mutation Operators}

SuMo implements the novel \textbf{SFR} (SafeMath Function Replacement) operator that targets the widely used SafeMath library. SafeMath\footnote{\url{https://docs.openzeppelin.com/contracts/2.x/api/math#SafeMath}} provides arithmetic functions that automatically revert the transaction in case of overflow. Developers often use them in place of the base arithmetic operators to increase code reliability. The SFR operator replaces each SafeMath function call with a different one, causing the calculation of the wrong value. 

\subsection{General Mutation Operators}
This section illustrates the general mutation operators implemented by SuMo, which are summarized in table \ref{tab:sumo.solidity.operators}. General mutation operators target traditional programming constructs that are not unique to the Solidity language. SuMo implements 19 general mutation operators, 4 of which are novel. Several operators were inspired by well-documented tools such as muJava\cite{Ma2005,Ma2014,Ma2016} and PIT (\url{https://pitest.org/quickstart/mutators/}), and adapted to the Solidity language. In particular, PIT's operators were selected because they focus on limiting the generation of redundant and equivalent mutants.

\subsubsection{Expression Mutation Operators}
The operators supported by Solidity are similar to the ones available in JavaScript. Binary mutation operators target binary expressions, which consist of two operands and one operator. Unary mutation operators target unary expressions, which consist of one operand and one operator. Lastly, assignment mutation operators target the assignment of a value to a variable, which can either be another variable or a literal.
\textbf{BOR} (Binary Operator Replacement): The BOR mutation operator targets all binary arithmetic, relational, conditional, bitwise and shift operators. In particular, BOR replaces a binary operator with a single binary operator of the same class to simulate typographical errors. \textbf{UORD} (Unary Operator Replacement or Deletion): The UORD mutation operator targets all unary arithmetic, bitwise and conditional operators. Unary operators are either removed or replaced with a different operator of the same class.
\textbf{AOR} (Assignment Operator Replacement): AOR replaces a compound assignment operator with a different one. AOR was designed following the same pattern as PIT's Math operator. This allows reducing the number of redundant mutants as well as the total number of mutants.  \textbf{ICM} (Increments Mirror): SuMo implements a modified version of the Increments Mirror operator. The usage of this operator for Solidity was first introduced by the Vertigo\cite{Honig2019} tool, which imported it from PIT. If a developer accidentally types \texttt{=-} instead of \texttt{-=}, the statement is interpreted in the wrong way. In fact, \texttt{=-} is an assignment followed by the unary -, while \texttt{-=} is a shortcut assignment that performs a decrement operation.

\subsubsection{Control Structures Mutation Operators}
Solidity supports most of the known control structures. These are: \texttt{if}, \texttt{else}, \texttt{while}, \texttt{do}, \texttt{for}, \texttt{break} and \texttt{continue}. The \texttt{return} statement was discussed in detail in section \ref{function.operators}. Solidity includes \texttt{try/catch} block for catching exceptions and reacting to the failure. In Solidity a faulty loop statement could execute the wrong number of iterations and also lead to \textit{out-of-gas} exception at run time. SuMo implements many mutation operators for control structures. In particular, we included the \textbf{BCRD} (Break and Continue Replacement and Deletion) operator, the \textbf{CBD} (Catch Block Deletion) operator, the \textbf{CSC} (Conditional Statement Change) operator, and finally the \textbf{LSC} (Loop Statement Change) operator. 

\subsubsection{Literal Mutation Operators}
Literal mutation operators replace hard-coded values within a contract with a different fixed value. In SuMo we included the
\textbf{BLR} (Boolean Literal Replacement) operator that simply replaces true with false and false with true. 
The \textbf{ILR} (Integer Literal Replacement) operator replaces any integer literal with a different value. In particular, each value is incremented and decremented by one.
The \textbf{HLR} (Hexadecimal Literal Replacement)  operator replaces a hexadecimal literal with a zero or non zero hexadecimal value.
The  \textbf{SLR} (String Literal Replacement) operator replaces any string literal with the empty string. 

\subsubsection{Types and Conversions Mutation Operators}
Types and conversions mutation operators target data types that are not unique to Solidity. The novel \textbf{ER} (Enum Replacement) operator implemented by SuMo helps to ensure the correct usage of Enums. Enums are broadly used in smart contract development, as they allow the creation of user-defined types. Each Enum can contain one or more members, where the first member represents the default value. If a developer is unaware of the default Enum behavior, it might lead to wrong assumptions about the value stored in a variable. It is also possible that an Enum variable is mistakenly assigned the wrong value, which can cause unexpected contract behavior. The ER operator swaps the first and second members of an Enum definition. This mutation forces the second member to become the default value of the Enum. It also replaces the member of an Enum assignment with a different member. This causes the affected Enum variable to assume a different value than intended. SuMo also implements the novel \textbf{ECS} (Explicit Conversion to Smaller type) operator that helps to avoid dangerous mistakes during explicit conversions. ECS targets explicit integer conversions to simulate truncation faults. In particular, it forces a conversion to the smallest integer type to cause the loss of higher-order bits. Similarly, it forces explicit byte conversions to a smaller type than intended. This can cause the right-most bits to be truncated.

\subsubsection{Overriding Mutation Operators}
Solidity's overriding mechanism allows child contracts to redefine the behavior of certain derived functions, modifiers, and constructors. The overriding mechanism must be enabled on each public or internal function with two distinct keywords: 
%\begin{itemize}
    %\item 
    A \texttt{virtual} function can be overridden by a derived contract;
    %\item
    An \texttt{override} function has been overridden in the current contract.
%\end{itemize}
Function modifiers and contract constructors can be overridden in the same manner, although they do not support the usage of the \texttt{super} keyword (or the contract name). SuMo implements several operators that target the overriding mechanism. The \textbf{ORFD} (Overridden Function Deletion), \textbf{SKD}  (Super Keyword Deletion) and \textbf{SKI} (Super Keyword Insertion) operators ensure the correct implementation and invocation of overridden methods. These operators were inspired by Deviant, which is the only other framework that implements overriding mutation operators. In addition, SuMo implements the \textbf{OMD} operator that targets the modifier overriding mechanism (section \ref{function.modifiers.operators}).

\subsubsection{Overloading Mutation Operators}
Function overloading allows a smart contract to implement multiple functions with the same name, but with a different number (or type) of parameters. The most problematic aspect of overloading is that it can increase human confusion, which is the primary source of bugs. If a function is heavily overloaded, a developer could forget which version of the method does what. Moreover, when many versions of the same method are available, it is more likely to call an unintended method. Overusing overloading can also lead to the implementation of trivial methods which are not needed. This can cause further confusion, as well as an increased amount of gas for contract deployment. The novel \textbf{OLFD} and \textbf{ACM} operators implemented by SuMo were designed to test different aspects of method overloading in Solidity. The \textbf{OLFD} (Overloaded Function Deletion) operator removes the declaration of each overloaded method within the contract, one at a time. This operator is useful in two ways. First, it allows identifying errors in the invocation of overloaded methods. If the mutant contract still works as expected without the deleted method, it means that such a method is not being called correctly. This can happen, for instance, when the developer invokes the wrong method version. Besides, OLFD allows to ensure coverage of overloaded methods. The test suite can only notice the mutation if the missing method is actually being called. The \textbf{ACM} (Argument Change of Overloaded Method Call) operator alters the invocation of overloaded methods within the contract. In particular, it swaps an overloaded method call with one different, existing overloaded method call. The mutated invocation must have different arguments to match a different overloaded method. This causes a different method version to be invoked.
\subsection{SuMo Implementation}
%\label{sec:architecture}

SuMo (\url{http://pros.unicam.it/sumo/}) 
has been implemented so to permit running a complete mutation testing process on a Solidity project. To this end, SuMo automatically generates and tests the mutants for a set of user-selected Solidity files.
SuMo implements 44 mutation operators that can be enabled and disabled through a CLI. It is also possible to exclude selected contracts, such as libraries and contract migrations, from the mutation process. These features allow the developers to easily customize the mutation testing process to their needs. When the testing process is started, SuMo uses the solidity-parser-antlr (\url{https://github.com/ConsenSys/solidity-parser-antlr}) for generating the AST (Abstract Syntax Tree) of each input SCUT (Smart Contract Under Test). The AST is then visited and mutated according to the rules specified by each enabled operator. Once the mutations have been applied, SuMo attempts to compile and test each mutant by running the provided test suite. Information about the mutation testing process, such as the generated stillborn mutants and the mutation score, is then saved to a final test report. A copy of each generated mutation is also saved in a file. This allows the tester to examine the survivors for excluding the equivalent mutants and enhancing the test suite.

\section{Validation}
\label{sec:validation}

To evaluate the effectiveness of our approach we selected two open-source dApps from GitHub, with associated test suites, and we mutated each project with SuMo. We chose applications showing some complexity and with different characteristics to have the possibility of targeting a wide range of Solidity features.\\
%\begin{itemize}
     %\item 
     1. \textbf{EtherCrowdfunding}\footnote{\url{https://github.com/giobart/EtherCrowdfunding}}: this is a dApp for organizing crowdfunding campaigns. It supports the creation of a new campaign, the donation of funds from the donors, and the reception of funds from the beneficiaries.\\
    %\item 
    2. \textbf{bionic-event-dapp}\footnote{\url{https://github.com/Mikearaya/bionic_event_dapp}}: this is an application for buying and selling event tickets. It supports the creation and management of events, as well as the purchase, validation, and transfer of tickets.
%\end{itemize}

\begin{table}
\setlength{\tabcolsep}{3.5pt}
\caption{Subject application metrics}
\label{tab:subject.dapp.metrics}
\centering
\begin{tabular}{lcccccc}
\toprule
  \textbf{dApp} & \textbf{\#Files} & \textbf{LOC}  & \textbf{\#Tests}  & \textbf{\makecell[c]{Test\\LOC}} & \textbf{\makecell[c]{Stmt\\Coverage}} & \textbf{\makecell[c]{Branch\\Coverage}}   \\
\midrule
 \makecell[l]{Ether\\Crowdfunding}   & 1 & 428 & 35 & 902 & 93.83 & 71.3\\
bionic-event  & 2 & 182 & 25 & 261 & 90 & 65\\
\bottomrule
\end{tabular}
\end{table}

\begin{table}
\caption{Mutated contracts of the subject applications}
\label{tab:mutated.contracts}
\centering
\begin{tabular}{ll}
\toprule
  \textbf{dApp} & \textbf{Mutated Contracts}\\
\midrule
\textbf{EtherCrowdfunding} & \makecell[l]{CrowdfundingCampaign.sol} \\  
\textbf{bionic-event}  &           \makecell[l]{Event.sol\\EventFactory.sol} \\
\bottomrule
\end{tabular}
\vspace{-0.6cm}
\end{table}

Table \ref{tab:subject.dapp.metrics} shows the metrics of each subject application. For each dApp, the table describes the name (Col. 1), number of mutated contracts (Col. 2), lines of code (Col. 3), test suite size (Col. 4), lines of code of the test suite (Col. 5), statement coverage (Col. 6) and branch coverage (Col. 7). The coverage values were calculated with the solidity-coverage (\url{https://github.com/sc-forks/solidity-coverage}) tool. Both applications reach almost full statement coverage values and relatively good branch coverage. Table \ref{tab:mutated.contracts} shows the contract files that were selected for mutation. All the meaningful contracts of each project were mutated with both general and Solidity-specific operators. 
Table \ref{tab:experimental.results} shows the results of the experiments carried out on the selected projects. For each dAPP, the table shows the number of generated mutants (Col. 2), stillborn mutants (Col. 3), equivalent mutants (Col. 4), and the mutation score (Col. 5). At the end of each test run, we performed a manual inspection to identify and remove the equivalent mutants generated by SuMo. The results show that the selected test suites cannot achieve high mutation scores. This confirms that test suites with good coverage values do not necessarily ensure the reliability of the contract code. In addition, it is possible to observe that the mutation score and code coverage are limitedly correlated. \newline
Our analysis revealed that some classes of mutants are less likely to be killed than others. The \textit{EED} mutations were regularly missed by tests. Since most mutations were easily reachable from the test suite, this means that events are often overlooked by testers. However, the emission of events (or lack thereof) can be extremely useful to ensure the correct behavior of the contract. The \textit{EHC} mutants survived most of the time. Most exception handling statements were commented out by EHC to prevent their execution. These results suggest a general lack of tests that attempt to make the condition check fail. Indeed, developers often neglect unhappy paths. This cannot ensure the proper functioning of the error handling mechanisms in the contract. Our results also show that \textit{AVR} mutants were rarely killed. This operator introduced many different types of fault in the SCUT, such as wrong address assignments, address balances, or event parameters. AVR was also useful to expose the lack of test coverage for a contract function. The self-destruct mechanism seems to be poorly tested as well. In particular, the alive \textit{SFD} mutants indicate that the existing tests never attempt to destroy the contract. This can be dangerous, as any fault in the self-destruct mechanism can prevent the contract from being deleted from the blockchain. The \textit{FVR} and \textit{VVR} mutants were often missed by tests as well. Most survivors contain a looser visibility keyword than the original contract. This result was predictable, as testers rarely check whether unauthorized accounts can call restricted functions. Mutations that swap a visibility keyword with a more restrictive one are also likely to be missed, especially if the semantic difference with the original contract is very subtle. It is safe to assume that function visibility is rarely targeted because defining dedicated test cases requires additional effort. However, analyzing the survivors can challenge the developer to understand if the contract functions and variables are associated with the correct visibility keyword. \textit{GVR} mutants were missed by the tests half of the time. Several mutants were killed because the usage of the wrong global variable can easily propagate and cause anomalous contract behavior. However, the survivors can expose a lack of test cases for ensuring the correct state of the contract. Operators that target function modifiers gave us mixed results. \textit{MOI} mutants achieved a low mutation score. A MOI mutant can survive if the tests manage to trigger the ``true" branch of the added modifier. For instance, the transaction might be issued from an account that passes the authorization check. However, it is also possible that the tests never attempt to call the mutated function. The \textit{MOD} mutants achieved a higher mutation score, but several mutations were still missed by the tests.
This further confirms a lack of tests that exercise the unhappy path, such as calling a function from an unauthorized account.
Lastly, \textit{MOR} mutants were always killed by tests. Replacing a modifier is a less subtle fault that can drastically change the behavior of the affected function. The selected case studies did not offer opportunities for evaluating the \textit{MOC} and \textit{OMD} operators. The \textit{SFR} operator generated few mutants, most of which were killed by tests. This was predictable as SFR causes arithmetic errors that can easily propagate through the contract. However, the survivor was useful to spot a lack of tests that ensure the correctness of the contract state after the execution of the affected method.\newline
To further validate the efficacy of SuMo we
enhanced the test suite of the \textit{bionic-event-dapp} project based on its survivors. Examining the live mutants allowed us to improve the existing test data and to design additional test cases. Table \ref{tab:bionic.event.enhanced} shows the results for the second mutation testing run. The mutation score has sensitively increased, which indicates the higher fault-detection capability of the new test suite.

\begin{table}
\caption{Experimental Results}
\label{tab:experimental.results}
\setlength{\tabcolsep}{3.5pt}
\centering
 \begin{tabular}{lccccc}
       \toprule
        \textbf{dApp} & \makecell[c]{\textbf{Mutation}\\\textbf{Operators}} & \makecell[c]{\textbf{Total}\\\textbf{Mutants}} & \makecell[c]{\textbf{Stillborn}\\\textbf{Mutants}} 
        & \makecell[c]{\textbf{Equivalent}\\\textbf{Mutants}}
         & \makecell[c]{\textbf{MS(\%)}}
        \\
        \midrule
          \multirow{2}*{\makecell[l]{\textbf{Ether}\\\textbf{Crowdfunding}}}
               & All   & 681  & 59 & 13 &  47,73 \\
                & Solidity & 401 & 38 & 7 &  28,93  \\
           \midrule
           \multirow{2}*{\textbf{bionic-event}}
               & All   & 189  & 29 & 34 & 58,7 \\
                & Solidity & 148 & 29 & 24 & 54,7  \\
            \bottomrule
    \end{tabular}
\end{table}

\begin{table}
\caption{Results for the bionic-event-dapp project after enhancing the test suite}
\label{tab:bionic.event.enhanced}
\setlength{\tabcolsep}{3.5pt}
\centering
    \begin{tabular}{cccccc}
       \toprule
     \makecell[c]{\textbf{Mutation}\\\textbf{Operators}} & \makecell[c]{\textbf{Total}\\\textbf{Mutants}} & \makecell[c]{\textbf{Stillborn}\\\textbf{Mutants}} 
        & \makecell[c]{\textbf{Equivalent}\\\textbf{Mutants}}
         & \makecell[c]{MS(\%)}
        \\
        \midrule
               All   & 172  & 29 & 32  & 86,84 \\
               Solidity & 136 & 29 & 22 & 85,05 \\
            \bottomrule
    \end{tabular}
    \vspace{-0.6cm}
\end{table}

\subsection{Threats to validity}
\label{sec:threats}

It is rather clear that, even though we got some encouraging results, the performed validation does not permit us to derive solid answers for the real effectiveness of the applied approach. There are several aspects that have to be deepened. The various choices in the definition of the strategy, for instance how to reduce stillborn mutants, have been mainly driven by our experience, and we did not have the opportunity to check their real efficacy. At the same time, performance and scalability aspects have to be considered, and checked, in order to get concrete indications on the applicability of the approach to real contexts. To summarize, the validation we performed permitted us to get some indications, but it is necessary to enlarge it both in relation to the considered dApps, and to other relevant qualitative aspects. 
\section{Related Works}
\label{sec:related}

Several existing tools implement a mutation testing approach for Ethereum smart contracts. Wu et al. propose \textbf{MuSC} \cite{Wu2019, Li2019}, the first mutation testing framework for the assessment of Ethereum smart contract test suites. The tool is open source and can be found on Github. 
This work proposes nine JavaScript-oriented operators and fifteen Solidity-specific mutation operators that can inject faults in the smart contract code. The proposed set of operators is moderately compact, which allows limiting the total number of generated mutants. However, it lacks several operators that can introduce severe faults or vulnerabilities into the contract, such as the ones that target function modifiers. 
While the mutation operators used by MuSC were designed starting from the Solidity documentation, the approach proposed by Andesta et al. \cite{Andesta2019} is based on the study of known bugs in Solidity smart contracts. This set of Solidity operators was included in the \textbf{Universal Mutator} tool. The core idea of this work is that designing operators capable of recreating real-world bugs can be an effective way for selecting good test cases. This led to the definition of 10 classes of operators that cover many types of smart contract vulnerabilities. The results show that the proposed set leads to the generation of a big amount of invalid and redundant mutants, while the rate of generation of equivalent mutants was not mentioned. The evaluation carried out by Andesta et al. shows that the proposed operators can successfully recreate 10 out of 15 buggy contracts. However, no actual mutation scores were provided by the authors.
Chapman \cite{Chapman2019} proposes \textbf{Deviant}, an open-source mutation testing tool that aims to help developers in delivering higher quality code. In this case, a broader set of 62 Solidity-specific mutation operators was designed according to the Solidity fault model.
However, it ends up generating a large number of mutants that require a big computational effort to be executed against the test suite. 
The \textbf{ContractMut} tool proposed by Hartel \& Schumi \cite{Hartel2019} implements general mutation operators derived from the minimum standard Mothra set, as well as four Solidity specific mutation operators. To define a more manageable set of operators and limit the number of generated mutants, the authors apply selective mutation techniques. In particular, the mutation operators were designed to only target the most severe vulnerabilities, with a focus on mistakes concerning access control. As a result, the percentage of stillborn mutants generated by this framework is rather small (15.8\%). The effectiveness of this approach was evaluated by producing replay tests derived from historic transaction data on the blockchain. Replay tests have a limited code coverage with respect to a real test suite, which would likely achieve higher mutation scores.

\section{Conclusions and Future Work}
\label{sec:conclusion}

This work presented a novel mutation testing strategy and a corresponding fully functional tool for Solidity smart contracts. SuMo implements a comprehensive set of mutation operators for simulating a wide range of traditional and Solidity-specific faults and vulnerabilities. The analysis of the Solidity documentation and of existing tools allowed to design 11 novel mutation operators, 7 of which target the unique features of Solidity. In particular, SuMo introduces mutation operators that target the overloading mechanism, which was never considered by related works. SuMo also introduces new operators for the contract constructor, function modifiers, cryptographic global functions, the SafeMath library, global blockchain variables, enums, return values, and explicit conversions. We have evaluated the effectiveness of SuMo by running the mutation testing process on two real-world DApps. The preliminary results show that the average mutation score, got by the test suites associated with the two dApps, is moderately low, especially for the case of Solidity-specific features. We were not able to assess all the novel operators, but some generated mutants survived the defined test suites for the two case studies. Further experiments are needed to assess if the surviving mutants can also lead to the identification of relevant faults. Indeed future work will mainly focus on extending the evaluation activity, and on providing a GUI for SuMo.

\bibliographystyle{IEEEtran}
\bibliography{biblio}%

\begin{thebibliography}{10}
\providecommand{\url}[1]{#1}
\csname url@samestyle\endcsname
\providecommand{\newblock}{\relax}
\providecommand{\bibinfo}[2]{#2}
\providecommand{\BIBentrySTDinterwordspacing}{\spaceskip=0pt\relax}
\providecommand{\BIBentryALTinterwordstretchfactor}{4}
\providecommand{\BIBentryALTinterwordspacing}{\spaceskip=\fontdimen2\font plus
\BIBentryALTinterwordstretchfactor\fontdimen3\font minus
  \fontdimen4\font\relax}
\providecommand{\BIBforeignlanguage}[2]{{%
\expandafter\ifx\csname l@#1\endcsname\relax
\typeout{** WARNING: IEEEtran.bst: No hyphenation pattern has been}%
\typeout{** loaded for the language `#1'. Using the pattern for}%
\typeout{** the default language instead.}%
\else
\language=\csname l@#1\endcsname
\fi
#2}}
\providecommand{\BIBdecl}{\relax}
\BIBdecl

\bibitem{Porru2017}
S.~Porru, A.~Pinna, M.~Marchesi, and R.~Tonelli, ``Blockchain-oriented software
  engineering: challenges and new directions,'' in \emph{39th ICSE
  Companion}.\hskip 1em plus 0.5em minus 0.4em\relax IEEE, 2017, pp. 169--171.

\bibitem{sac20-chorchain}
F.~Corradini, A.~Marcelletti, A.~Morichetta, A.~Polini, B.~Re, and F.~Tiezzi,
  ``Engineering trustable choreography-based systems using blockchain,'' in
  \emph{ACM SAC}, 2020, p. 1470–1479.

\bibitem{Zheng2017}
Z.~Zheng, S.~Xie, H.~Dai, X.~Chen, and H.~Wang, ``An overview of blockchain
  technology: Architecture, consensus, and future trends,'' in
  \emph{International Congress on Big Data)}.\hskip 1em plus 0.5em minus
  0.4em\relax {IEEE}, 2017, pp. 557--564.

\bibitem{Destefanis2018}
G.~Destefanis, M.~Marchesi, M.~Ortu, R.~Tonelli, A.~Bracciali, and R.~Hierons,
  ``Smart contracts vulnerabilities: a call for blockchain software
  engineering?'' in \emph{International Workshop on Blockchain Oriented
  Software Engineering ({IWBOSE})}.\hskip 1em plus 0.5em minus 0.4em\relax
  {IEEE}, 2018, pp. 19--25.

\bibitem{Mehar2019}
M.~I. Mehar, C.~L. Shier, A.~Giambattista, E.~Gong, G.~Fletcher, R.~Sanayhie,
  H.~M. Kim, and M.~Laskowski, ``Understanding a revolutionary and flawed grand
  experiment in blockchain,'' \emph{Journal of Cases on Information
  Technology}, vol.~21, no.~1, pp. 19--32, 2019.

\bibitem{Miraz2020}
M.~H. Miraz and M.~Ali, ``Blockchain enabled smart contract based applications:
  Deficiencies with the software development life cycle models,''
  \emph{Information Systems eJournal}, 2020.

\bibitem{Zou2019}
W.~Zou, D.~Lo, P.~S. Kochhar, X.-B.~D. Le, X.~Xia, Y.~Feng, Z.~Chen, and B.~Xu,
  ``Smart contract development: Challenges and opportunities,'' \emph{{IEEE}
  Transactions on Software Engineering}, 2019.

\bibitem{Inozemtseva2014}
L.~Inozemtseva and R.~Holmes, ``Coverage is not strongly correlated with test
  suite effectiveness,'' in \emph{IEEE/ACM ICSE 2014}, 2014.

\bibitem{Tengeri2016}
D.~Tengeri, L.~Vidacs, A.~Beszedes, J.~Jasz, G.~Balogh, B.~Vancsics, and
  T.~Gyimothy, ``Relating code coverage, mutation score and test suite
  reducibility to defect density,'' in \emph{Int. Conf. on Software Testing,
  Verification and Validation ({ICSTW})}.\hskip 1em plus 0.5em minus
  0.4em\relax {IEEE}, 2016, pp. 174--179.

\bibitem{wood2019ethereum}
\BIBentryALTinterwordspacing
G.~Wood, ``Ethereum: A secure decentralised generalised transaction ledger
  byzantium version,'' 2019. [Online]. Available:
  \url{https://ethereum.github.io/yellowpaper/paper.pdf}
\BIBentrySTDinterwordspacing

\bibitem{ethereumWhitepaper}
V.~Buterin, ``Ethereum whitep.'' \url{https://ethereum.org/whitepaper/}, 2013.

\bibitem{SolidityDoc}
``Solidity doc.'' \url{https://solidity.readthedocs.io/en/latest/index.html}.

\bibitem{Atzei2017}
N.~Atzei, M.~Bartoletti, and T.~Cimoli, ``A survey of attacks on ethereum smart
  contracts,'' in \emph{LNCS}.\hskip 1em plus 0.5em minus 0.4em\relax Springer,
  2017, vol. 10204, pp. 164--186.

\bibitem{Dika2018}
A.~{Dika} and M.~{Nowostawski}, ``Security vulnerabilities in ethereum smart
  contracts,'' in \emph{2018 IEEE iThings and IEEE GreenCom and IEEE CPSCom and
  IEEE SmartData}, 2018, pp. 955--962.

\bibitem{Andesta2019}
E.~Andesta, F.~Faghih, and M.~Fooladgar, ``Testing smart contracts gets
  smarter,'' \emph{CoRR}, vol. abs/1912.04780, 2019.

\bibitem{Chapman2019}
P.~Chapman, D.~Xu, L.~Deng, and Y.~Xiong, ``Deviant: A mutation testing tool
  for solidity smart contracts.''\hskip 1em plus 0.5em minus 0.4em\relax IEEE,
  2019, pp. 319--324.

\bibitem{Honig2019}
J.~J. Honig, M.~H. Everts, and M.~Huisman, ``Practical mutation testing for
  smart contracts,'' in \emph{LNCS}.\hskip 1em plus 0.5em minus 0.4em\relax
  Springer, 2019, vol. 11737, pp. 289--303.

\bibitem{Hartel2019}
P.~H. Hartel and R.~Schumi, ``Mutation testing of smart contracts at scale,''
  in \emph{14th International Conference, TAP@STAF 2020}, ser. LNCS, vol.
  12165.\hskip 1em plus 0.5em minus 0.4em\relax Springer, 2020, pp. 23--42.

\bibitem{Wu2019}
H.~Wu, X.~Wang, J.~Xu, W.~Zou, L.~Zhang, and Z.~Chen, ``{Mutation Testing for
  Ethereum Smart Contract},'' vol. abs/1908.03707, 2019.

\bibitem{Ma2005}
Y.-S. Ma, J.~Offutt, and Y.~R. Kwon, ``{MuJava}: an automated class mutation
  system,'' \emph{Software Testing, Verification and Reliability}, vol.~15,
  no.~2, pp. 97--133, 2005.

\bibitem{Ma2014}
Y.-S. Ma and J.~Offutt, ``Description of class mutation mutation operators for
  java,'' \emph{Electronics and Telecom. Research Institute, Korea}, 2005.

\bibitem{Ma2016}
Y.~Ma and J.~Offutt, ``Description of mujava’s method-level mutation
  operators,'' \emph{Update}, 2016.

\bibitem{Li2019}
Z.~Li, H.~Wu, J.~Xu, X.~Wang, L.~Zhang, and Z.~Chen, ``{MuSC: A tool for
  mutation testing of ethereum smart contract},'' \emph{ASE}, pp. 1198--1201,
  2019.

\end{thebibliography}

\end{document}